\documentstyle[anta, twoside]{article}

\input{psfig.sty}

\def\Journal#1#2#3#4{(#1) {#2} {\bf #3}, #4}

\def\AAp{\em Astron. Astrophys.}

\def\ApJ{\em Astrophys.~J.}

\def\ApJSS{\em Astrophys.~J. Suppl.}

\newcommand{\RM}{{\rm RM}}
\newcommand{\pheins}{\phantom{1}}
\newcommand{\phelf}{\phantom{11}}
\newcommand{\phdrei}{\phantom{111}}
\newcommand{\phminus}{\phantom{$-$}}
\begin{document}

\markboth{X. H. Sun \& J. L. Han}{Structure Function Studies for Turbulent Interstellar
Medium}

\thispagestyle{plain}
\setcounter{page}{25}

\title{Structure Function Studies for Turbulent Interstellar Medium}

\author{X. H. Sun and J. L. Han}

\address{National Astronomical Observatories, Chinese Academy of Sciences,
Beijing, 100012, China}


\maketitle

\abstract{We study structure functions of rotation
measures in the Canadian Galactic Plane Survey (CGPS) region
and the North Galactic Pole (NGP) to extract the interstellar
medium (ISM) fluctuation information. The CGPS data are divided into
three longitude intervals: $82^\circ<l<96^\circ$ (CGPS1),
$115^\circ<l<130^\circ$ (CGPS2) and $130^\circ<l<146^\circ$ (CGPS3).
The structure functions of all three regions have large
uncertainties when the angular separation is smaller than
$\delta\theta\approx 1^\circ$. A power law can fit the structure function
well for $\delta\theta>1^\circ$. The power law indices get smaller from
CGPS1 to CGPS3 and the amplitudes decrease. The variations of the
large-scale field and the electron density have only negligible
effects on the structure function and thus cannot account for the
changes, indicating that the turbulent properties of the Galactic ISM
are intrinsically longitude-dependent. The Kolmogorov-like fluctuation
spectrum of the electron density or the magnetic field should produce a
power law structure function with an index of 5/3 or 2/3, neither of
which is consistent with our results of small indices in the three
sub-CGPS regions. For the NGP region, the structure function
is flat, showing that the rotation measures are mostly intrinsic
to the extragalactic sources, and the ISM is very random in that part
of our Galaxy. It is obvious that the ISM fluctuation is
latitude-dependent when comparing the results in the NGP region and the
CGPS regions.}

\section{Introduction}

The electron density irregularities have been studied either by
refractive scintillation or by diffractive scintillation phenomena as
systematically summarized by Armstrong et al.~(1995). The
fluctuation spectrum from scales of $\sim10^6$~m to scales of
$\sim10^{13}$~m can be fitted by a single power law with a
power index of $\sim3.7$, i.e. the Kolmogorov spectrum. This spectrum
can be extended up to $\sim 10^{18}$~m. The property of magnetic field
fluctuations is not yet clear although they are more important than
the electron density fluctuations in some circumstances such as in
the solar wind (Armstrong et al. 1995).

The structure function of rotation measures (RMs) contains information
on electron density and magnetic field fluctuations (Simonetti et al. 1984).
More data are available now allowing a better determination of the
structure function and a careful comparison of the turbulent properties
of different regions. In this contribution, we show the RM structure functions
of extragalactic sources in the Canadian Galactic Plane Survey
(Brown et al. 2003, CGPS hereafter) and briefly discuss the results.
The detailed report of this work will be given elsewhere.

\section{The Theoretical Structure Function}
The RM structure function ($D_{\RM}$), if assumed as a stationary random
process, can be written as
\begin{equation}
D_{\RM}({\bf \delta\theta})=
\langle[\RM({\bf \theta_0})-\RM({\bf \theta_0+\delta\theta})]^2\rangle\ ,
\end{equation}
where $\delta\theta$ is the angular separation of two sources for RMs in
degree throughout this paper, and $\langle\cdots\rangle$ denotes the ensemble
average over any $\theta_0$ so that the result is independent of $\theta_0$.
Both the electron density ($n_e$) and the magnetic field (${\bf B}$)
are assumed to consist of uniform backgrounds ($n_0$ and ${\bf B_0}$)
and Gaussian fluctuations ($\delta n$ and $\delta {\bf B}$) with
averages of zero, in the form of
\begin{eqnarray}
n_e&=&n_0+\delta n\ ,\nonumber\\
{\bf B}&=&{\bf B_0}+{\bf \delta B}\ .\nonumber
\end{eqnarray}
The exact spectra of the electron density and magnetic field
fluctuations are not known, so it is straightforward to take the power
law form as
\begin{eqnarray}
P_{\delta
n}(q)&=&\frac{C_n^2}{[l_{on}^{-2}+q^2]^{\alpha/2}}\ ,\nonumber\\
P_{\delta B}(q)&=&\frac{C_B^2}{[l_{oB}^{-2}+q^2]^{\beta/2}}\ ,\nonumber
\end{eqnarray}
where $q$ is the wave number, $C_n^2$ and $C_B^2$ are the fluctuation
intensities, $l_{on}$ and $l_{oB}$ are the outer scales, and $\alpha$
and $\beta$ are the power indices. The structure function can be
reduced to the form below when $l_{on}=l_{oB}=l_o=1/q_o$ (Minter 1995)
\begin{eqnarray}
D_{\RM}(\delta\theta)&=&\Psi_{\RM}(\delta\theta)+
2C_{\RM}^2B_{0z}^2C_n^2f(\alpha)L^{\alpha-1}\delta\theta^{\alpha-2}+
2C_{\RM}^2n_0^2C_B^2g(\beta)L^{\beta-1}\delta\theta^{\beta-2}\nonumber\\
&& \nonumber\\
&&+2C_{\RM}^2C_n^2C_B^2h(\alpha,\beta)L^{(\alpha+\beta)/2-1}
q_o^{3-(\alpha+\beta)/2}
\delta\theta^{(\alpha+\beta)/2-2}\ , \nonumber
\end{eqnarray}
where $\Psi_{\RM}(\delta\theta)$ is the geometrical term attributed
to the variation of the large-scale electron density and the magnetic
field, $L$ is the path length, $f(\alpha)$, $g(\beta)$ and
$h(\alpha,\beta)$ are constants containing $\Gamma$-functions.
As $\delta\theta\ll1$~rad, the geometrical term can be approximated as
\begin{displaymath}
\Psi_{\RM}(\delta\theta)\approx C_{\RM}^2n_0^2B_0^2L^2\delta\theta^2
\approx 3\times10^{-4}\langle \RM\rangle^2\delta\theta^2\ ,
\end{displaymath}
here $\delta\theta$ is in degree. If both the electron density and the
magnetic field fluctuations follow a Kolmogorov spectrum, the structure
function has a simple form as below,
\begin{displaymath}
D_{\RM}(\delta\theta)=\Psi_{\RM}(\delta\theta)+C\delta\theta^{5/3}\ ,
\end{displaymath}
where $C$ is a constant related to the intensities in the regular
and fluctuating components.

\section{Analysis of Data}
In principle, two methods can be employed to extract the fluctuation
information of the ISM from RMs: the autocorrelation function and the
structure function. The autocorrelation function is the Fourier
transform of the power spectrum density according to the {\it
Wiener-Khinchin theorem}, so the direct way is to derive the
autocorrelation function. But this can cause problems when the data
are irregularly spaced (Spangler et al. 1989). Regardless of the data
sampling, it is easy to obtain the structure function, however. So we
will take the structure function approach below.

Given a sample of RM data, the logarithm of angular separations of RM
pairs are set to bins with equal lags. In each bin the differences of
RM pairs are squared and then averaged to obtain the structure function.
The standard deviations are taken as the uncertainty of each function
value.

Around each source we can derive the local RM average and dispersion.
If a RM value is different from the average by three times larger than
the dispersion, it is defined as ``anomalous''. We will illustrate by
simulation that such an anomalous RM can significantly affect the structure
function.

Let the RMs distribute in the region
$0^\circ\leq x\leq2^\circ$ and $0^\circ\leq y\leq2^\circ$ uniformly
with grid intervals of 0.1$^\circ$. Two samples of random RM values
following Gaussian distributions with an average of zero and
a dispersion $\sigma$ of 30 and 50 rad m$^{-2}$, respectively, are generated
with the Monte-Carlo method. Note that the average value does not
affect the results. Theoretically the structure function should
be flat with an amplitude $2\sigma^2$ and an index 0. The structure
functions from simulations are displayed in Fig. 1. We then fitted the
structure functions with $D_{\RM}(\delta\theta)=A\delta\theta^\alpha$
and found that the results are consistent with theoretical expectations.

\begin{figure}[!htbp]
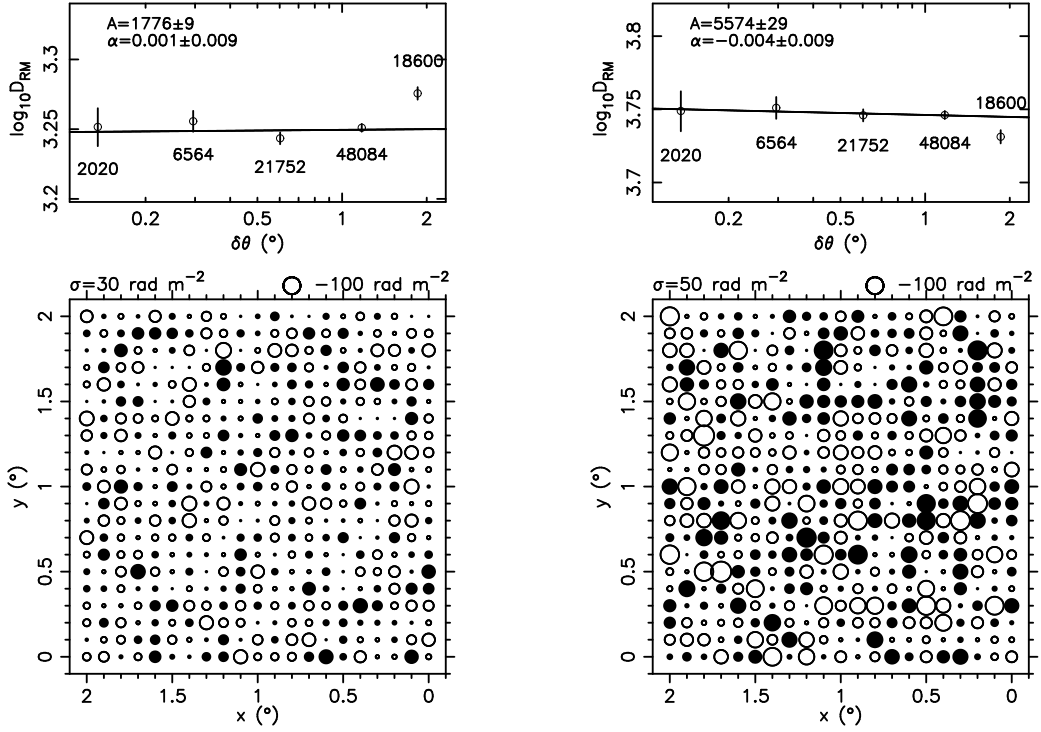

\begin{minipage}{0.5\textwidth}
\centerline{\psfig{figure=XHSun_Fig1_1.ps,width=6cm,angle=-90}}
\end{minipage}\hfill
\begin{minipage}{0.5\textwidth}
\centerline{\psfig{figure=XHSun_Fig1_2.ps,width=6cm,angle=-90}}
\end{minipage}\hfill
\caption{Simulated RM distributions are shown in the lower panels,
with filled and open circles representing positive and negative values,
respectively. The sizes of symbols are proportional to the square-root
of RM values. The structure functions are plotted in the upper panels
with the number of pairs in each bin marked.}
\end{figure}

\begin{figure}[!htbp]
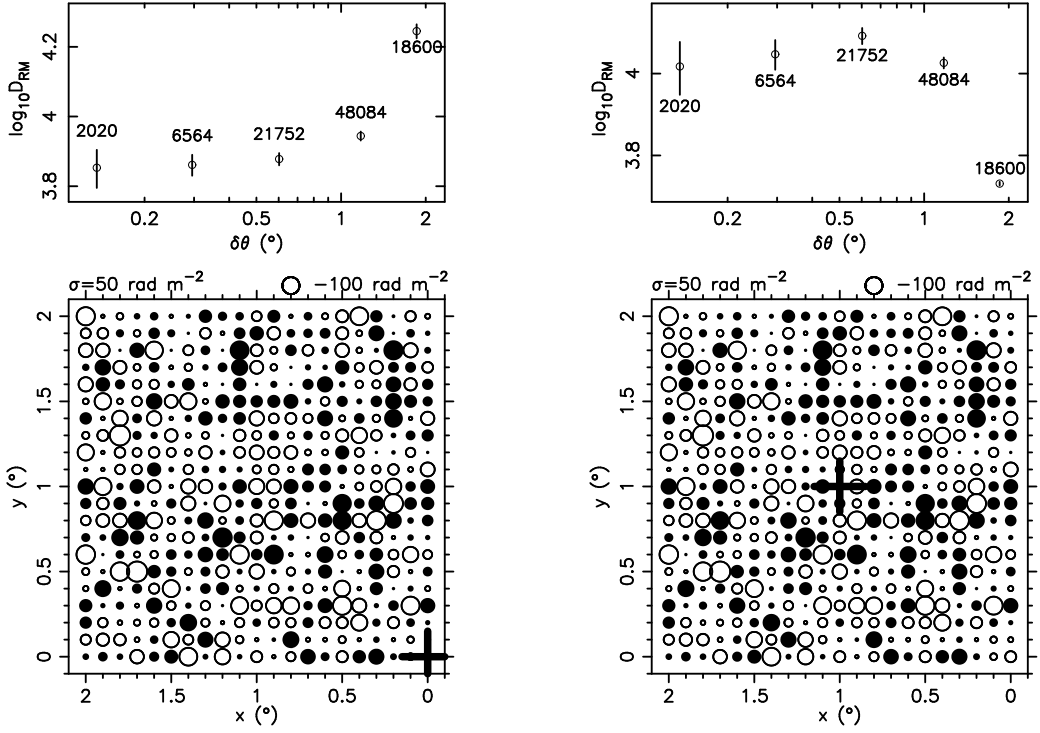

\begin{minipage}{0.5\textwidth}
\centerline{\psfig{figure=XHSun_Fig2_1.ps,width=6cm,angle=-90}}
\end{minipage}\hfill
\begin{minipage}{0.5\textwidth}
\centerline{\psfig{figure=XHSun_Fig2_2.ps,width=6cm,angle=-90}}
\end{minipage}\\[5mm]
\caption{Same as Fig. 1 for $\sigma=50$~rad m$^{-2}$ but with
an anomalous RM of 1000 rad m$^{-2}$ in the corner and in the center
as indicated by a cross.}
\end{figure}

As is obvious in Fig.~2, an anomalous RM value (1000 rad m$^{-2}$,
indicated by a cross) can distort the real structure function to a
different extent. With regard to actual RM data, such a RM can be
either due to an enhanced local medium or it is intrinsic to the
extragalactic source. One should pick up such RMs for further
identification. Below we will only show the results of real observed RM
data after discarding these anomalous RMs, as identifications have not
yet been made.

We took the RMs of the extragalactic sources (EGSs) in the CGPS region
but excluded sources containing separate or unresolved components
(marked by a, b and c in the catalog of Brown et al. 2003).
The data are divided into three longitude intervals:
$82^\circ<l<96^\circ$ (CGPS1), $115^\circ<l<130^\circ$ (CGPS2) and
$130^\circ<l<146^\circ$ (CGPS3). We also collected the RMs of EGSs in
the North Galactic Pole (NGP) region with latitudes larger than
70$^\circ$. Then we calculated the structure functions for these four
regions, as shown in Figs.~3, 4, 5, and 6. The fitting results using
$D_{\RM}(\delta\theta)=A\delta\theta^{\alpha}$ are listed
in Table~1, where Column~1 refers to the region, Columns 2 and 3
to the average and dispersion of the RMs. The amplitude and power index
of each power law structure function are listed in Columns 4 and 5.

\begin{figure}[!htbp]
\begin{minipage}{0.5\textwidth}
\centerline{\psfig{figure=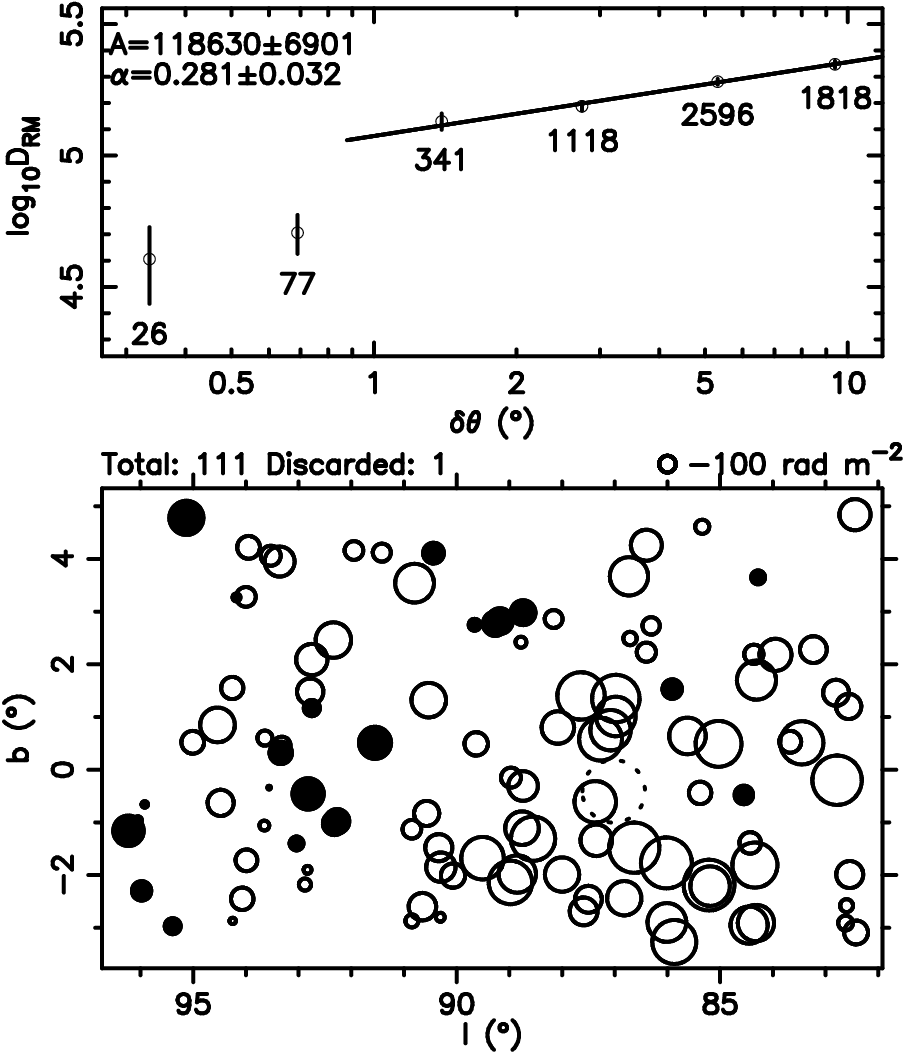,width=7cm,angle=-90}}
\end{minipage}%
\begin{minipage}{0.5\textwidth}
\centerline{\psfig{figure=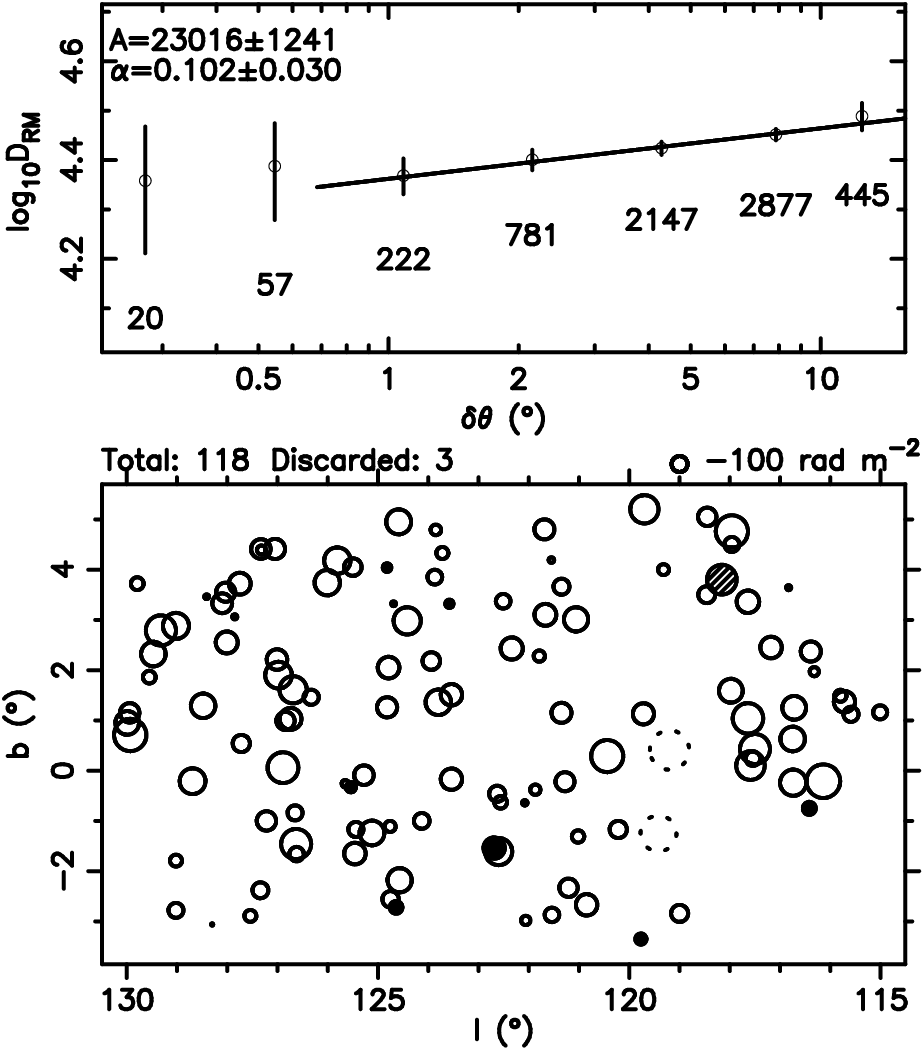,width=7cm,angle=-90}}
\end{minipage}
\begin{minipage}{0.5\textwidth}
\caption{The RM distribution in the CGPS1 region (lower panel)
and the calculated structure function (upper panel). The symbols
are plotted as in Fig.~1, but anomalous RMs are plotted as
hatched circles when positive or dotted circles when negative. }
\end{minipage}%
\begin{minipage}{0.5\textwidth}
\caption{Same as Fig.~3 but for the CGPS2 region.}
\end{minipage}
\end{figure}

\begin{figure}[!htbp]
\begin{minipage}{0.5\textwidth}
\centerline{\psfig{figure=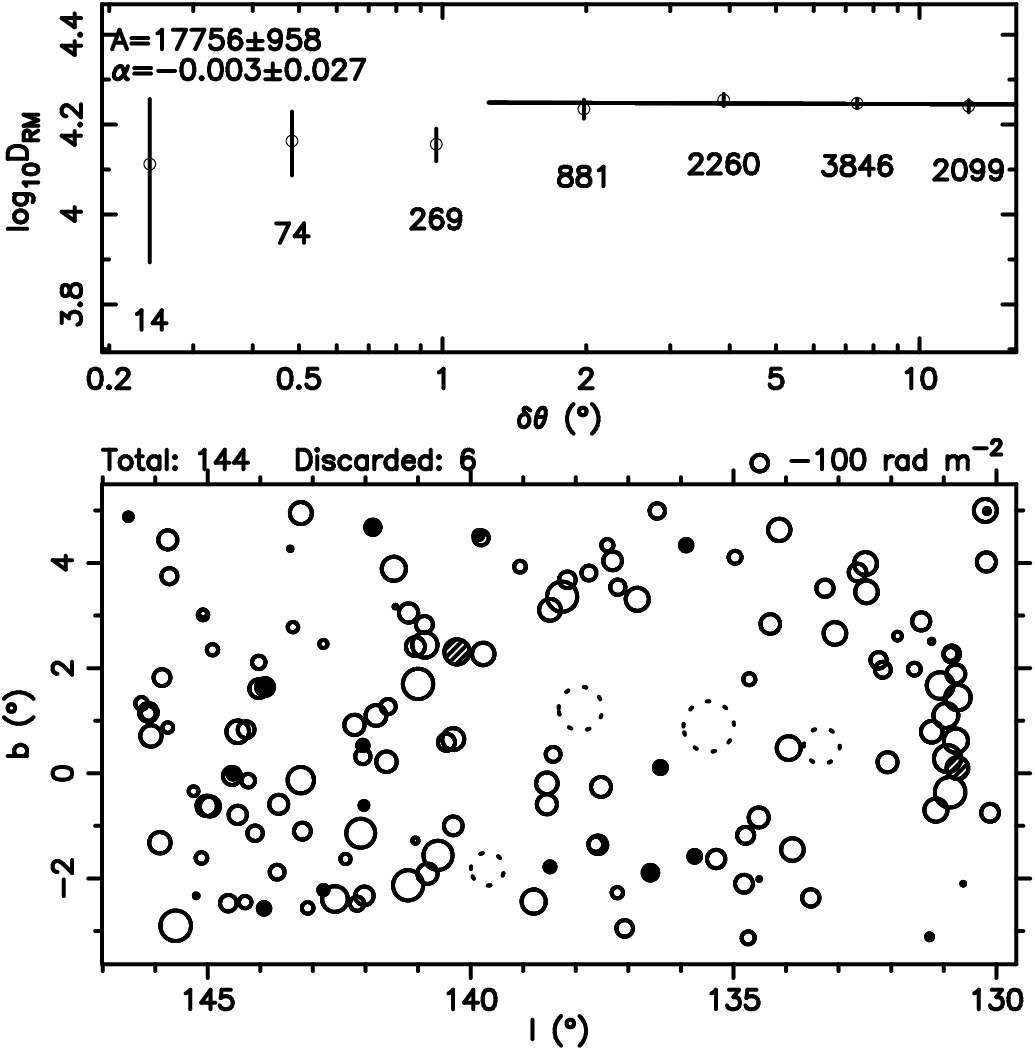,width=7cm,angle=-90}}
\caption{Same as Fig.~3 but for the CGPS3 region.}
\end{minipage}%
\begin{minipage}{0.5\textwidth}
\centerline{\psfig{figure=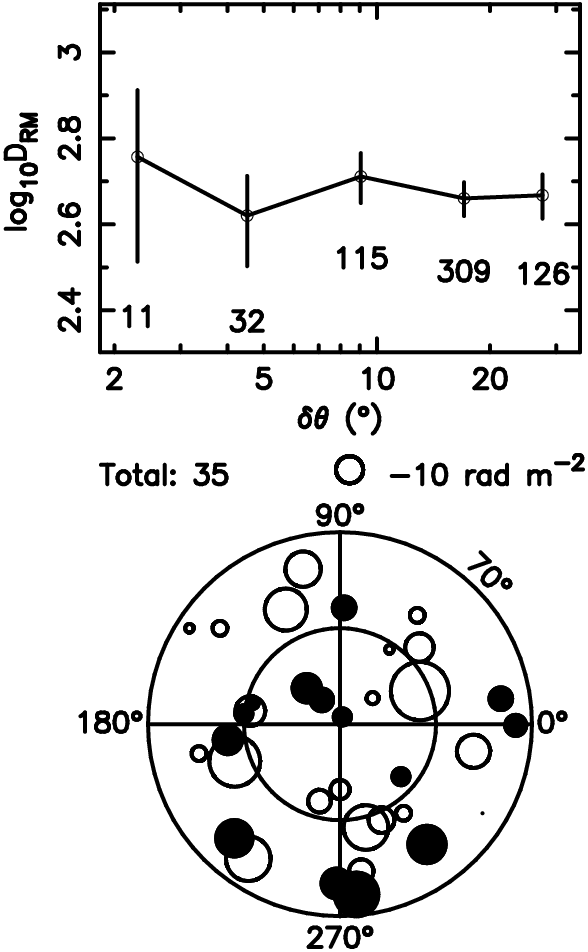,width=6cm,angle=-90}}
\caption{Same as Fig.~3 but for the NGP region.}
\end{minipage}
\end{figure}

\begin{table}
\caption{Fitting parameters of the structure functions in the CGPS regions
  and the NGP region.}
\begin{center}
\begin{tabular}{lccccc}\noalign{\medskip}\hline\noalign{\smallskip}
Region &Range &$\langle \RM\rangle$&$\sigma_{\RM}$&$A$&$\alpha$\\
\noalign{\smallskip}\hline\noalign{\smallskip}
CGPS1 &$\pheins 82\fdg 42<l< \pheins 96\fdg 23$
   &$-$237 & 306 & 118630$\pm$6901  &\phminus 0.28\pheins$\pm$0.03\pheins \\
CGPS2 &$115\fdg 01<l<129\fdg 99$  &$-$147 & 117 &  \pheins 23016$\pm$1241
   &\phminus 0.10\pheins$\pm$0.03\pheins \\
CGPS3 &$130\fdg 12<l<146\fdg 51$  &$-$106 & \pheins 94  &\pheins 17756$\pm$\pheins 958
   &$-$0.003$\pm$0.027\\
North Galactic Pole &($b>70^\circ$)   &\phelf $-$3  &\pheins 15  &\phdrei 548$\pm$\pheins 215
   &$-$0.06\pheins $\pm$0.12\pheins \\
\hline
\end{tabular}
\end{center}
\end{table}

\section{Discussion}
The electron density fluctuations are better known than the magnetic
field fluctuations. Armstrong et al. (1995) have shown that the
electron density fluctuation spectrum is Kolmogorov-like, but the
magnetic field fluctuation remains mysterious and totally unclear as
mentioned above. We know that many physical processes in the ISM affect
the electron density and the magnetic field at the same time, such as
supernova explosions. The magnetic field is often assumed to be frozen into
the interstellar medium. So we can assume that there is no distinct
discrepancy between the fluctuations of the electron density and the
magnetic field. In this case the structure function can be written
as $D_{\RM}(\delta\theta)=\Psi(\delta\theta)+C\delta\theta^{\alpha-2}$
where $C$ is a constant.

In the three sub-CGPS regions, the average rotation measures
$\langle \RM \rangle=-$237, $-$147 and $-$106 rad m$^{-2}$,
respectively, corresponding to the geometrical term contributions
$\Psi_{\RM}(\delta\theta)=17\delta\theta^2$, $6\delta\theta^2$ and
$3\delta\theta^2$, which are definitely negligible when compared to the
obtained amplitudes ($A$) of the structure functions in these regions.
In fact, the indices from our results are much smaller than 2, which
also means that the geometrical term plays a minor role in the structure
function. Therefore the structure function can be simplified as
$D_{\RM}(\delta\theta)\approx C \delta\theta^{\alpha-2}$.

The structure function $D_{\RM}(\delta\theta)\approx C
\delta\theta^{5/3}$ holds for 3D Kolmogorov fluctuations and
$D_{\RM}(\delta\theta)\approx C \delta\theta^{2/3}$ for 2D Kolmogorov
fluctuations (Minter \& Spangler 1996). {\it It is clear that the
Kolmogorov spectrum cannot account for our results.} Actually, the
indices of the structure functions in the three sub-CGPS regions are
very small, close to zero. Let us try to discuss the nearly flat
structure functions we got.

From the results (Table 1) it is evident that the amplitudes of the
structure function $A$ decrease from CGPS1 to CGPS3. This is probably
caused by the extent of the ISM in our Galaxy. From our simulation,
we know that $A\sim 2\sigma_{RM}^2$, where $\sigma_{RM}$ is the
dispersion of RM along the line of sight. Assuming that there are many
fluctuation cells with typical scale $l$ and a RM dispersion
$\sigma_{l}$ along the line of sight, the total RM dispersion can be
represented by $\sigma_{RM}^2=\frac{L}{l}\sigma_l^2$; where
$L$ is the path length to the edge of our Galaxy. Because $L$ decreases from
CGPS1 to CGPS3, it is understandable that the fluctuation intensity
gets smaller.

Due to the disk structure of the Galactic electron distribution and the
Sun's location at 8 kpc distance from the Galactic center, the RMs of
EGSs near the pole region are little affected by the Galactic interstellar
medium, so that the structure function for the EGSs in the pole region
should be very flat. Our result in Fig.~6 and the fitting parameters in
Table~1 are consistent with Simonetti et al. (1984) and strengthen the fact
that the RMs of EGSs in the pole region are very random, indeed mainly
of intrinsic origin. We can also see that the fluctuation intensity in the
NGP region is much smaller than that in the CGPS region, indicating a
trend of a latitude-dependence of the ISM turbulence.

\section{Conclusion}
We have used the structure function to study ISM fluctuations.
Simulation shows that an anomalous RM can significantly influence the
structure function, so we suggest that these large RMs should be
carefully checked before performing a structure function analysis, or
otherwise should be used with caution. The structure functions
of the RMs of EGSs in the CGPS region have been calculated.
The structure functions at $\delta\theta>1^\circ$ can be fitted by a
power law. The power indices are all nearly zero which cannot
be interpreted by Kolmogorov-like ISM fluctuations. The fluctuation
intensity is longitude-dependent. The flat structure function in the
NGP region shows that the RMs in this region are almost intrinsic to
EGSs.

\section*{Acknowledgments}
We would like to thank the LOC and SOC for the nice meeting in Turkey.
This work is partially supported by the National Natural Science
Foundation of China (19903003 and 10025313) and the National Key Basic
Research Science Foundation of China (G19990754) as well as the
partner group of MPIfR at NAOC.

\section*{References}\noindent

\references
Armstrong J. W., Rickett B. J., Spangler S. R.
  \Journal{1995}{\ApJ}{443}{209}.

Brown J. C., Taylor A. R., Jackel B. J.
  \Journal{2003}{\ApJSS}{145}{213}.

Minter A. H. (1995) {\em Ph.D. Thesis}, University of Iowa.

Minter A. H., Spangler S. R. \Journal{1996}{\ApJ}{458}{194}.

Simonetti J. H., Cordes J. M., Spangler S. R. \Journal{1984}{\ApJ}{284}{126}.

Spangler S. R., Fanti R., Gregorini L., Padrielli L. \Journal{1989}{\AAp}{209}{315}.

\end{document}